\documentclass[a4paper,fleqn]{cas-dc}

\usepackage[numbers]{natbib}

\def\tsc#1{\csdef{#1}{\textsc{\lowercase{#1}}\xspace}}
\tsc{WGM}
\tsc{QE}
\tsc{EP}
\tsc{PMS}
\tsc{BEC}
\tsc{DE}
\begin{document}
\let\WriteBookmarks\relax
\def\floatpagepagefraction{1}
\def\textpagefraction{.001}
\shorttitle{Call for establishing benchmark science and engineering. }
\shortauthors{Prof. Dr. Jianfeng Zhan}

\title [mode = title]{Call for establishing benchmark science and engineering.}

%{This is a specimen $a_b$ title}  
\tnotemark[1,2]

%\tnotetext[1]{This document is the results of the %research
%   project funded by the National Science %Foundation.}

%\tnotetext[2]{The second title footnote which is a longer text matter
%   to fill through the whole text width and overflow into
%   another line in the footnotes area of the first page.}

\author[1,3]{Jianfeng Zhan}
%[type=editor-in-chief,
%                        auid=000,bioid=1,
%                        prefix=Professor,
%                        role=Researcher,
%                        orcid=0000-0001-7511-2910]
\cormark[1]
\fnmark[1]
\ead{zhanjianfeng@ict.ac.cn or zhanjianfeng.benchcouncil@gmail.com}
\ead[url]{www.benchcouncil.org/zjf.html}

%\credit{Conceptualization of this study, Methodology, Software}

\address[1]{Institute of Computing Technology, Chinese Academy of Sciences and BenchCouncil}

%\author[2,4]{Han Theh Thanh}[style=chinese]

%\author[2,3]{CV Rajagopal}[%
%   role=Co-ordinator,
%   suffix=Jr,
%\ead{cvr3@sayahna.org}
%\ead[URL]{www.sayahna.org}

%\credit{Data curation, Writing - Original draft preparation}

%\address[2]{Sayahna Foundation, Jagathy, Trivandrum 695014, India}

%\author%
%[1,3]
%{Rishi T.}
%\cormark[2]
%\fnmark[1,3]
%\ead{rishi@stmdocs.in}
%\ead[URL]{www.stmdocs.in}

%\address[3]{STM Document Engineering Pvt Ltd., Mepukada,
%    Malayinkil, Trivandrum 695571, India}

%\cortext[cor2]{Principal corresponding author}
%\fntext[fn1]{This is the first author footnote. but is common to %third
%  author as well.}
%\fntext[fn2]{Another author footnote, this is a very long footnote and
%  it should be a really long footnote. But this footnote is not yet
%  sufficiently long enough to make two lines of footnote text.}

%\nonumnote{This note has no numbers. In this work we demonstrate $a_b$
%  the formation Y\_1 of a new type of polariton on the interface
%  between a cuprous oxide slab and a polystyrene micro-sphere placed
%  on the slab.
%  }

\begin{abstract}
Currently, there is no consistent benchmarking across multi-disciplines. Even no previous work tries to relate different categories of benchmarks in multi-disciplines. This article investigates the origin and evolution of the benchmark term. Five categories of benchmarks are summarized, including measurement standards, standardized data sets with defined properties, representative workloads, representative data sets, and best practices, which widely exist in multi-disciplines. I believe there are two pressing challenges in growing this discipline: establishing consistent benchmarking across multi-disciplines and developing meta-benchmark to measure the benchmarks themselves. I propose establishing benchmark science and engineering; one of the primary goals is to set up a standard benchmark hierarchy across multi-disciplines.  It is the right time to launch a multi-disciplinary benchmark, standard, and evaluation journal, TBench, to communicate the state-of-the-art and state-of-the-practice of benchmark science and engineering.   

\end{abstract}

%\includegraphics{figs/grabs.pdf}

%\item Research highlights item 1
%\item Research highlights item 2
%\item Research highlights item 3

\maketitle

\section{The origin and evolution of the benchmark term}\label{origin_evoluation}

%It is common practice  Olympic
%sprinters, for instance, may compare themselves against a time benchmark or against a
%close competitor. Beating the time benchmark or beating the competitor allows them

Benchmarking is common practice in all industries, and indeed in many areas of life~\cite{performance_evaluation_finance}. For example, an Olympic
sprinter or fund manager or IT product manager may compare themselves against a benchmark or 
a close competitor to evaluate their performance.
Unfortunately, the benchmark term independently evolves in multi-disciplines and has related but different implications. 
This section investigates the origin and evolution of the benchmark concept.

%in multi-disciplines.

%, to benchmark or compare performance.
%and then defines what benchmarks are in the big picture in the next section. %in this section.

I find that the modern benchmark concept (close to its current definition)  first appeared in measurement science~\cite{stevens1946theory} in the form of  bench mark (two words separated by a space). For example, in geodesy, a bench mark is a mark whose height, relative to datum, has
been determined by leveling---the operation to measure differences in height between established points relative to a datum~\cite{zairi1996origins}. Later, this concept is extended into multi-disciplines. 

In the computer discipline, one of the earliest benchmarking effort~\cite{lewis1985evolution} dated back to 1962 in the Auerbach Corporation's Standard EDP Reports. Joslin defined this benchmarking effort as "a routine used to determine the speed performance of a computer system."~\cite{lewis1985evolution}. The reports included reporting performance data using typical benchmark tasks---many basic functions---but based on the vendor's published data without stipulating that the benchmark must run on the system under test. Around 1965, Joslin~\cite{joslin1965evaluation} stated that the most important question in computer evaluation should be "how long will it take this system to process my workload (my computer application)?". This exploring methodology produced the concepts of workload modeling, application benchmark, synthetic benchmarks,  and standard benchmark, which are still used nowadays~\cite{lewis1985evolution}. These concepts seem abstract, not directly related to the bench mark concept, though having some connections. The primary reason may be that the computer is a new thing at that time. 

The followings are simple explanations of these concepts. Workload modeling is selecting a representative sample set of programs from the entire real workloads~\cite{lewis1985evolution}, which is a critical factor ensuring the benchmark quality.  
An application benchmark is a mix of programs to be run on several different computer configurations to obtain comparative performance in terms of handling the specific applications~\cite{joslin1965evaluation}. Because of the difficulty (cost) of porting real applications across different systems, in 1969, Bucholz~\cite{buchholz1969synthetic} argued a greater degree of abstraction---a synthetic benchmark to imitate the real application---is necessary to make comparisons across different systems practical. The rising costs of synthetic benchmarks motivated the standardization of benchmarks. In 1976, a group of government and industry personals was formed to ascertain the possibility of a standard benchmark library~\cite{conti1978findings}, which was the first try in this regard.

As a general term, in the 1987 edition of the Oxford Reference Dictionary, the benchmark is defined as a surveyor's mark indicating a point in a line of levels, a standard or point of reference~\cite{zairi1996origins}. The editors obviously did not consider the benchmark concept that appeared in the computer discipline, but their benchmark definition is similar to that in geodesy we referred at the beginning of this section; Zairi et al.~\cite{zairi1996origins} thought this definition is the beginning of today's use of the word benchmark in the management discipline.

In the management discipline,  the Xerox Corporation
was the pioneer of benchmarking~\cite{zairi1996origins}: its roots began in 1979, evaluated itself externally through this process which
became known as competitive benchmarking. This benchmarking research and practice~\cite{zairi1996origins} encompassed an in-depth, ongoing
study of best competitors, including detailed reverse engineering of competitor products, technology processes, what they achieved and how they did it,  and a tear-down analysis of operating capabilities and features of competing products.   This benchmarking practice is very similar to the computer discipline's benchmark-driven performance engineering in terms of the principle. The latter tries to disclose the root causes of the performance bottlenecks of and optimize the computer systems considering the specific workloads.

Gradually, benchmarking was 
extended as a strategic quality tool to all aspects of the business and 
progressively integrated into the management process~\cite{zairi1996origins}. 
In this context, Zairi et al.~\cite{zairi1996origins} defined it as the continuous process of measuring products, services, and processes against the industry best practices that lead to superior performance.

\section{Five categories of benchmarks}\label{five_benchmark}

This section investigates five categories of benchmarks in multi-disciplines. My intention is not to provide a consistent or unified benchmark definition. Instead, I try to reveal the essence of the benchmarks in five different scenarios. I leave the discussion of consistent benchmarking in the following two sections.

\begin{figure*}
\centering
\includegraphics[scale=0.2]{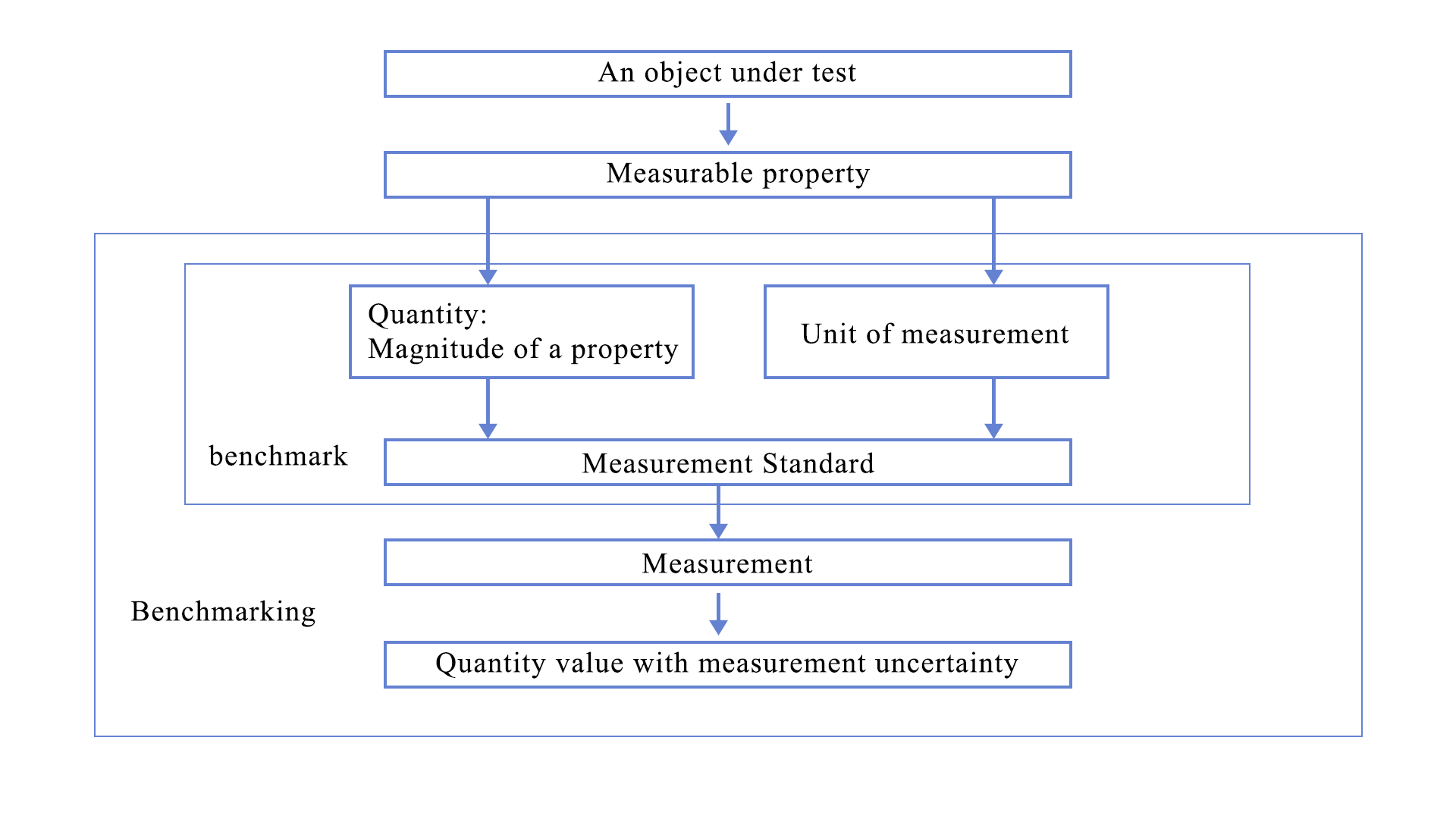}
\caption{The interpretation of the first category of the benchmark from the perspective of metrology~\cite{bipm2012international,kacker2021quantity}.}
\label{FIG:1}
\end{figure*}

The first category of the benchmark is a measurement standard.  In the computer discipline, the Linpack benchmark is of this category, which is widely used to report the performance of a high-performance computer.  I provide an interpretation of this category from the perspective of metrology. The Joint Committee for Guides in Metrology (JCGM)~\cite{bipm2012international} defines a measurement standard as a realization of the definition of a quantity, with stated value and associated measurement uncertainty, used as a reference.  As shown in Figure~\ref{FIG:1}, a benchmark realizes the definition of a quantity, the unit of measurement, the measurement methodology, and the reference implementation with stated measurement uncertainty. A quantity is a measurable property of the object under measurement, like length, energy, etc. Benchmarking covers two phases: the design and implementation of the benchmark and measuring the object's properties with the benchmark.
%is the process of using the benchmark to 

% we discuss three categories of benchmarks. The first one is of a measurement standard. As shown in Figure~\ref{FIG:1}, 

The second one is the representative workloads that run on the systems under measurement. The application benchmarks or synthetic benchmarks in the computer discipline, discussed in Section~\ref{origin_evoluation}, are of this category. They provide the design input to the system design and implementations.  They do not necessarily meet the stringent definition of measurement standards, but they are also used to evaluate systems.  For example, in the computer discipline, many deep learning workloads  (algorithms) are random with poor repeatability~\cite{tang2021aibench,jiang2021hpc}.  Deep learning is a kind of artificial intelligence (AI) workload.   However, they are representative workloads that can not be overlooked in the system design and implementation. 

\begin{figure*}
	\centering
		\includegraphics[scale=0.2]{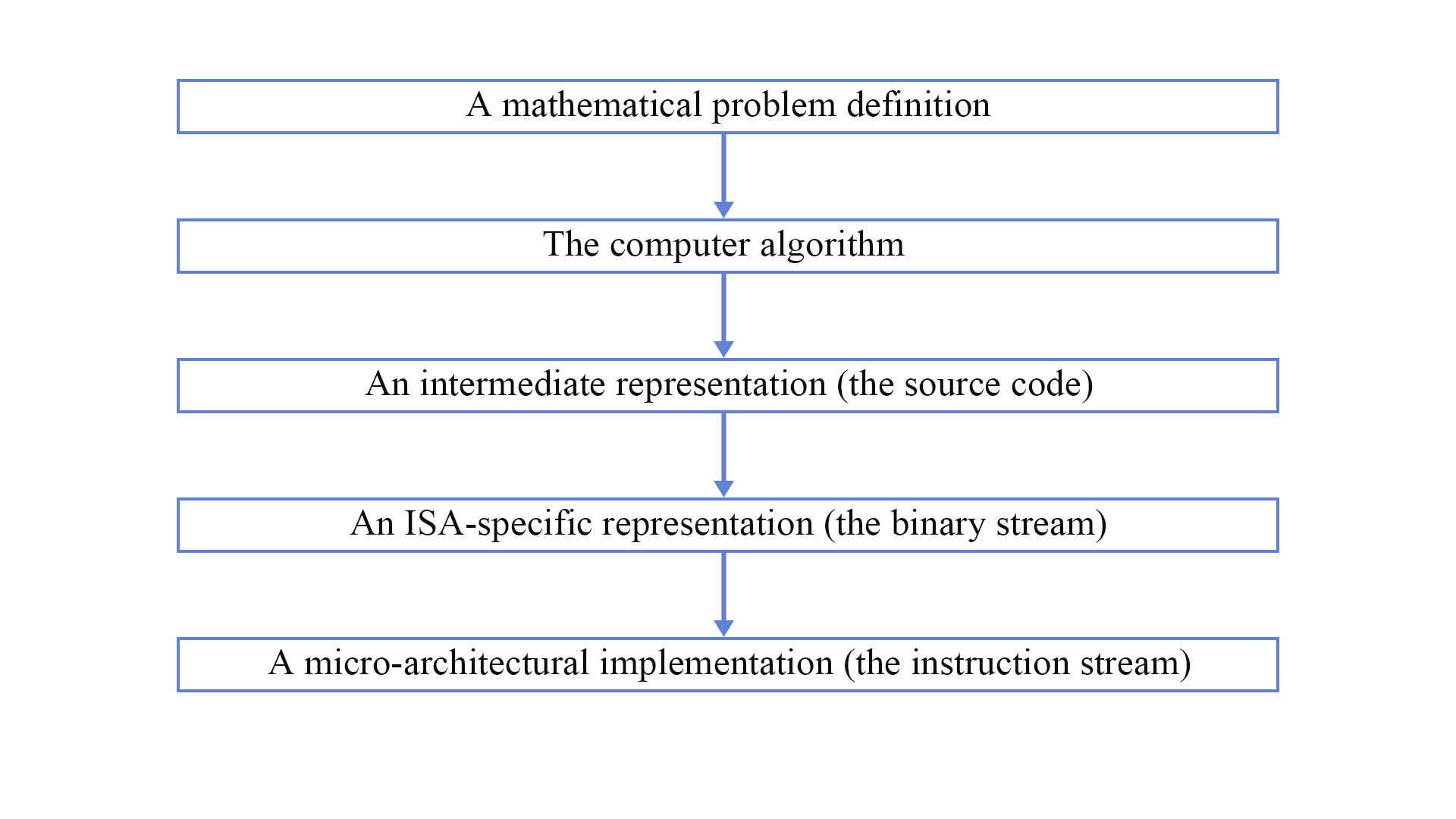}
	\caption{In the computer discipline, a representative workload,  the second category of the benchmarks, is hierarchically defined. From top to down is a mathematical problem definition, an algorithm, an intermediate representation, an ISA-specific representation, a micro-architectural representation. The lower level has more state space. State-of-the practice only analyzes a micro-architectural representation, which is only a subspace or even a point at a high-dimension space~\cite{wang2021wpc}. This hierarchy definition can be extended to other disciplines.}
	\label{FIG:2}
\end{figure*}

Generally speaking, the first category of the benchmarks is selected from the second category according to more strict criteria. Figure~\ref{FIG:2} explores how to define the representative workloads in the computer discipline. There is increasing freedom from a mathematical problem definition to an algorithm, an intermediate representation, An ISA-specific representation (ISA is short for instruction set architecture), and a micro-architecture representation. Section~\ref{challenges} will further discuss this challenge.

%A benchmark has a dual property. On the one hand, a benchmark is a measurement standard （property）.

%which is familiar to information technology or even business communities. On the other hand, 

%The second one is a it is a definition of a problem or even a challenge independent of its underlying implementations. The latter case emphasizes the definition of a problem or a challenge, though we can often use the implementation of a benchmark for measurement.

The third is a standardized data set that represents real-world data science problem~\cite{automl2021git}, with defined properties, some of which have ground truth. ImageNet~\cite{deng2009imagenet} (deep learning benchmark) and MIMIC-III~\cite{johnson2016mimic} (critical care benchmark) are typical examples. 
 The benchmark of this category is often used to measure against different algorithms. The state-of-the-art algorithm implementation plus the data set usually constitutes the benchmark of the second category. 

The fourth is a representative data set, used as a reference. For example, a financial benchmark is an index (statistical measure), calculated from a representative set of underlying
data, is used as a reference for financial instruments or contracts~\cite{iosco2013finben}. Well-known financial benchmarks
include the London Interbank Offered Rate (Libor) and the Euro Interbank Offered Rate ~\cite{iosco2013finben}.

The fifth is the industry best practices in different domains.  Benchmarking is the continuous process of searching the industry best practices that lead to superior performance and measuring products, services, and processes against th\-em~\cite{zairi1996origins}. The Xerox Corporation
pioneered and enhanced this benchmarking process.

%~\cite{zairi1996origins}

%Benchmarking is used to measure performance using a specific indicator (cost per unit of measure, productivity per unit of measure, cycle time of x per unit of measure or defects per unit of measure) resulting in a metric of performance that is then compared to others.[1]

%Also referred to as "best practice benchmarking" or "process benchmarking", this process is used in management in which organizations evaluate various aspects of their processes in relation to best-practice companies' processes, usually within a peer group defined for the purposes of comparison. This then allows organizations to develop plans on how to make improvements or adapt specific best practices, usually with the aim of increasing some aspect of performance. Benchmarking may be a one-off event, but is often treated as a continuous process in which organizations continually seek to improve their practices.

%\section{The critical roles of benchmarking in multi-disciplines}

%We discuss their critical roles in multi-discipline

\section{The challenges}~\label{challenges}

As I elaborate in Section~\ref{five_benchmark}, the five categories of benchmarks have a closely connected. However, currently, there is no consistent benchmarking across multi-disciplines. Even no previous work tries to relate those five categories of benchmarks in multi-disciplines. The metrology science paves a foundation for this direction. However, they mainly focus on classical quantities like length, time, and power. Significantly different from those classical 
quantities, the properties of the objects in the computer,  management, or finance disciplines are greatly affected by its mathematical problem definition and concrete implementation, which raises a serious challenge. 

Different observation angles may distort the observable properties. For example, shown in Figure~\ref{FIG:2}, the quantity value of a computer workload is greatly affected by mathematical problem definitions, concrete algorithms,  different ISA and micro-architecture implementations. 

I further take the first category of benchmarks as an example to demonstrate the importance of tackling this challenge. Measuring “Quantum Supremacy” against the classical supercomputer is a fundamental issue.  Google’s “Quantum Supremacy” declaration in 2019~\cite{arute2019quantum} stated that the Syca\-more superconductive quantum computer (200 s) is over a billion times faster than the state-of-the-practice Summit system in 2016~\cite{wells2016announcing} (10,000 years) in the task of measuring and simulating one million samples. However, in 2021, a group of scientists and engineers declared, on the Sunway Supercomputer~\cite{liu2021closing}, they reduced the classical simulation sampling time of Google Sycamore to 304 s, from the previously claimed 10,000 years through both algorithmic and architecture innovations.  

The speed up---the ratio of the quantity values of two different kinds of systems---definitely will change wildly in the future. Understanding the benchmark very well under a hierarchy like that defined in Figure~\ref{FIG:2} is a priority before correctly interpreting the implication of the speed up, or else it will mislead the scientific community. The situation may become much complex in the other disciplines, as a clear hierarchy definition is also a luxury.  Establishing consistent benchmarking across multi-disciplines is very challenging.

%must be corrected understood with is greatly affected by the definition of a mathematical problem number only makes sense with 

The other challenge is how to measure the benchmarks themselves. Previous work has a preliminary discussion on this issue. For example, in the computer discipline, the characteristics of a (good) benchmark, i.e., representative~\cite{lewis1985evolution,pan2005finding}, relevance, reproducible, fair, verifiable, repeatable, and economical
 are discussed in ~\cite{v2015build, huppler2009art}. However, most of those properties are subjective. We need a meta-benchmark to evaluate those benchmarks. 
 
 I take the representative characteristic as an example; the current theory and practice can not convince the community that this topic is seriously treated.  From the perspective of mathematics, it is necessary to establish a mathematical foundation and consider the meaning of representative in a high dimension space.  Unfortunately, in practice, the benchmarking methodology seems ad-hoc. For example, it is reported that there are 6.8 million apps in the leading app stores~\cite{app2021git}.  How does the community infer the mobile phone market's representative workloads (and benchmarks)?  
 
\section{The proposal}~\label{proposal}

I believe that it is necessary to establish benchmark science and engineering; one of the goals is to set up standard benchmark hierarchy across multi-disciplines.  There are two reasons. First, there is a natural hierarchy in different categories of benchmarks. As we discussed in Section~\ref{five_benchmark}, the first benchmark category is selected from the second category according to more strict criteria. Second, through this hierarchy, we can tackle the challenge of the rising cost of benchmarking. For example, we can put more resources on the primary benchmarks while relating the other benchmarks to the primary benchmarks through traceability.

\begin{figure*}
	\centering
		\includegraphics[scale=0.2]{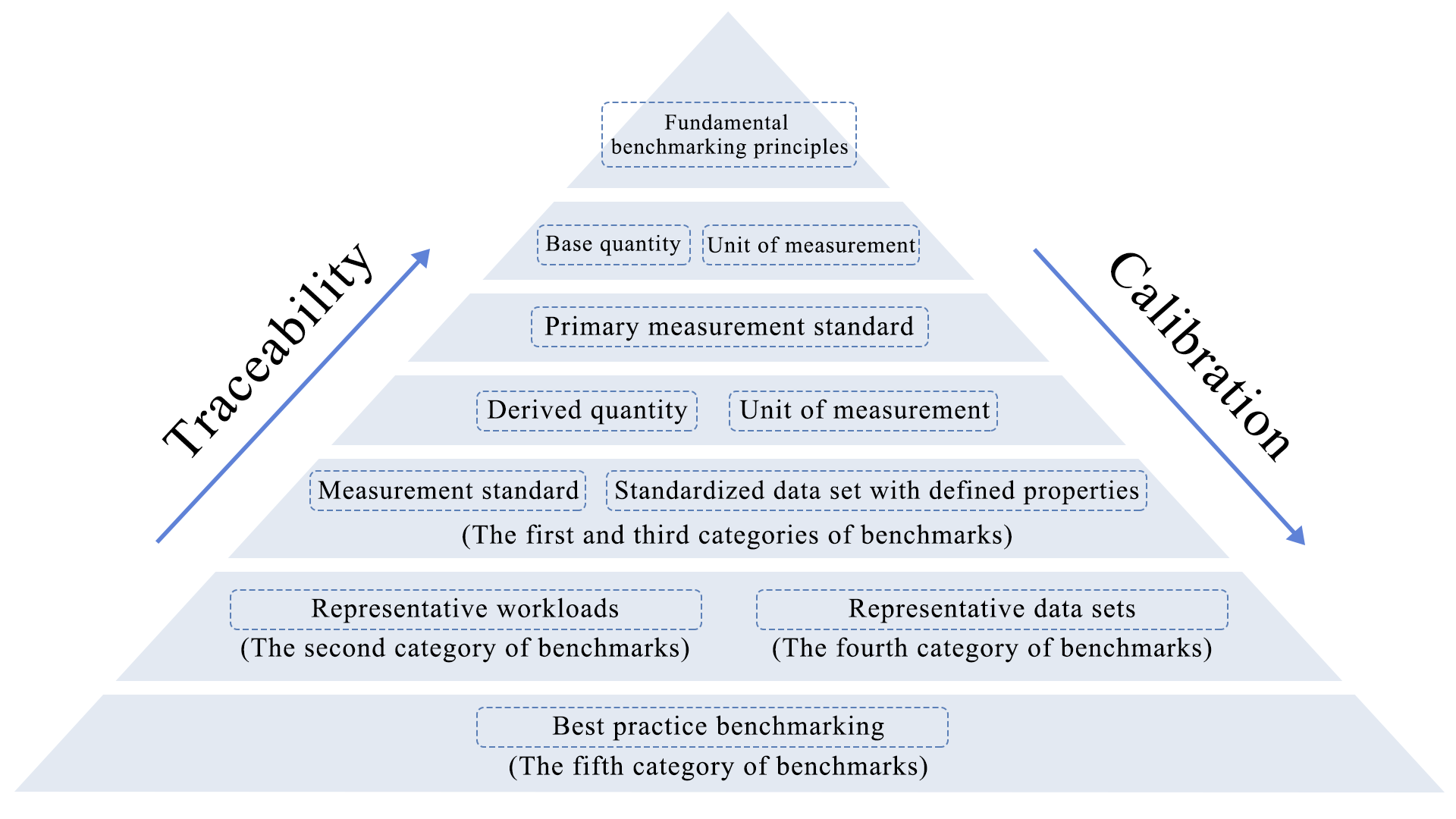}
	\caption{The standard benchmark hierarchy proposal.}
	\label{FIG:3}
\end{figure*}

Figure~\ref{FIG:3} is my proposal. The most important is to keep benchmarking consistently, and the following measures will help achieve the target: (1) the unified definition of base quantity and units of measurement; (2) the realization of quantities and units of measurement with different accuracy (and hence cost)  levels; (3)  the traceability and calibration across the standard benchmark hierarchy.  Traceability~\cite{bipm2012international} is a property of a measurement result whereby the result can be related to a reference through a documented unbroken chain of calibrations, each contributing to the measurement uncertainty.

  At the first tier, the international community needs to define the fundamental benchmarking principle and realize the base quantity, unit of measurement, primary measurement standard, which is the reference of all other benchmarks. The second tier is the first and second categories of the benchmarks. They will reuse the definitions and realizations of base quantity and unit of measurement from the first tier. Meanwhile, the definition and realization of derived quantity and unit of measurement are necessary. 

The third tier is the second and fourth categories of the benchmarks. The community often needs to revisit and ponder the mathematical or data problem definitions to provide state-of-the-art and state-of-the-practice implementations.  The fourth tier is the fifth category of the benchmarks. As it searches for the best practice, keeping an eye on the advancement of all hierarchies is necessary. 

%Traceability and calibrations are two fundamental processes to keep benchmarking consistently. 

 %is that the aim of the benchmark is to measure the  property 

%For example, for the properties of representative, we need a mathematics foundation. For the 

%explore the open issues and challenges in these vital areas

%The subjective definition of a good benchmark.

%The lack of a unified benchmark definition.

%the definition of unit of measurement. The traceablility (taking into account of the cost).

%The disciplined verification.  

%The lack of tool. 

%The quatum benchmark example. 

%The intrinsic properties. in computer science, are mainly observed from a specific-micro architecture.

%we need new tool

%The discussion of representativeness.

%we will provide a framework

%\section{The proposal}~\label{proposal}

\section{TBench: the venue for benchmark science and engineering}

I think it is the right time to launch a new journal, BenchCouncil Transactions on Benchmarks, Standards, and Evaluations (in short, TBench). It will provide a venue to communicate and tackle the challenges mentioned above as there is no multidisciplinary and interdisciplinary journal on this area.  I only noticed in the management discipline a closely related journal named Benchmarking: An International Journal.

The vital importance of a new journal is to guarantee that high-quality submissions receive high-quality reviews promptly.  According to the past experiences in the other reputable journals  and conferences in the computer discipline, which is my primary background, I  have some considerations.  

In the computer discipline, a journal paper often can not receive consistent and timely reviews compared with other top-tier conferences.  For example, different associate editors invite reviewers from uncertain sources to handle papers with large deviations.  Instead, a program committee meeting provides comparatively consistent reviews at a top-tier conference.  

Another issue is the significant delay. Overall, the average turnaround of handling a paper is from three months to a year. Some journals reject most submissions at the disposal of a staff who does not understand its content to speed up the process and reduce the external review load. That will harm our community for two reasons. First, the value of peer review is to provide constructive feedback, which is the stone of our scientific community. Second, it will result in the abuse of editor rights. 
The last issue is most journals adopt a single-blind review, which prevents fair review.

To resolve the above issues, I enact the following plans. (1) Consistent and reliable reviews. In addition to about thirty founding editors or editors, similar to the program committee member of a conference, we will invite approximately 30 associate editors (Junior researchers with Ph.D. degrees).  The associate editor is similar to the external review committee member of a  conference.  A team of founding editors, editors, and associate editors will provide the basis for consistent and reliable reviews.  

%If you have recommendations on associate editors, please let us know.

(2) Fast-track peer review.  The editor-in-chief (EIC) will read each paper's abstract and introduction. Suppose the team thinks this is a high-quality paper with high impact potential. In that case, they will invite three editors to have a timely review, including possible remote discussion, and make a final decision within three weeks.  The team will ask one editor and two associate editors to review the other papers. Overall, the team will finish one round of decisions within one month.  

(3) A double-blind review process. One member of the EIC team without conflict of interest (COI) is responsible for checking COIs, while the other EIC and editor, who do not know the authors' identities, make a final decision. Each published article is reviewed by a minimum of three independent reviewers using a double-blind peer-review process. The identities of the reviewers are not known to the authors, and the reviewers also do not know the identities of the authors.

%Finally, we explain why we launch a new non-profit organization---BenchCouncil (International Open Benchmark Council), to promote these activities. 

%This is the right venue discussing how to tackle the above challenge.
%\section{ What is a benchmark?}

\section{Acknowledgments}
I am very grateful to many persons' contributions to TBench, especially  Prof. Dr. Tony Hey for discussing the TBench plan, Dr. Lei Wang for discussing and proofreading this article,  Mr. Shaopeng Dai for compiling the references, Mr. Qian He for drawing the figures, Mr. Zhengxin Yang for discussing the metrology related work, Ms. Chitra Krishnamoorthy,  Ms. Divyaa Veluswamy, and other KeAI and Elsevier staffs for publishing TBench. Without all of you, launching TBench is impossible. 

%the PengCheng Laboratory for hardware support. We also thank Shaomeng Cao, Xuhui Shao, Yongheng Liu, Changsong Liu, and Jingfei Qiu for technical support in using those systems. 

%\appendix
%\section{My Appendix}
%Appendix sections are coded under \verb+\appendix+.

%\verb+\printcredits+ command is used after appendix sections to list 
%author credit taxonomy contribution roles tagged using %\verb+\credit+ 
%in frontmatter.

\printcredits

\bibliographystyle{cas-model2-names}

\bibliography{cas-refs}

\begin{thebibliography}{23}
\expandafter\ifx\csname natexlab\endcsname\relax\def\natexlab#1{#1}\fi
\providecommand{\url}[1]{\texttt{#1}}
\providecommand{\href}[2]{#2}
\providecommand{\path}[1]{#1}
\providecommand{\DOIprefix}{doi:}
\providecommand{\ArXivprefix}{arXiv:}
\providecommand{\URLprefix}{URL: }
\providecommand{\Pubmedprefix}{pmid:}
\providecommand{\doi}[1]{\href{http://dx.doi.org/#1}{\path{#1}}}
\providecommand{\Pubmed}[1]{\href{pmid:#1}{\path{#1}}}
\providecommand{\bibinfo}[2]{#2}
\ifx\xfnm\relax \def\xfnm[#1]{\unskip,\space#1}\fi
%Type = Article
\bibitem[{Arute et~al.(2019)Arute, Arya, Babbush, Bacon, Bardin, Barends,
  Biswas, Boixo, Brandao, Buell et~al.}]{arute2019quantum}
\bibinfo{author}{Arute, F.}, \bibinfo{author}{Arya, K.},
  \bibinfo{author}{Babbush, R.}, \bibinfo{author}{Bacon, D.},
  \bibinfo{author}{Bardin, J.C.}, \bibinfo{author}{Barends, R.},
  \bibinfo{author}{Biswas, R.}, \bibinfo{author}{Boixo, S.},
  \bibinfo{author}{Brandao, F.G.}, \bibinfo{author}{Buell, D.A.}, et~al.,
  \bibinfo{year}{2019}.
\newblock \bibinfo{title}{Quantum supremacy using a programmable
  superconducting processor}.
\newblock \bibinfo{journal}{Nature} \bibinfo{volume}{574},
  \bibinfo{pages}{505--510}.
%Type = Article
\bibitem[{BIPM et~al.(2012)BIPM, IFCC, IUPAC and ISO}]{bipm2012international}
\bibinfo{author}{BIPM, I.}, \bibinfo{author}{IFCC, I.}, \bibinfo{author}{IUPAC,
  I.}, \bibinfo{author}{ISO, O.}, \bibinfo{year}{2012}.
\newblock \bibinfo{title}{The international vocabulary of metrology—basic and
  general concepts and associated terms ({VIM}), 3rd edn. jcgm 200: 2012}.
\newblock \bibinfo{journal}{JCGM (Joint Committee for Guides in Metrology)} .
%Type = Article
\bibitem[{Buchholz(1969)}]{buchholz1969synthetic}
\bibinfo{author}{Buchholz, W.}, \bibinfo{year}{1969}.
\newblock \bibinfo{title}{A synthetic job for measuring system performance}.
\newblock \bibinfo{journal}{IBM Systems Journal} \bibinfo{volume}{8},
  \bibinfo{pages}{309--318}.
%Type = Inproceedings
\bibitem[{Clare(2014)}]{performance_evaluation_finance}
\bibinfo{author}{Clare, A.}, \bibinfo{year}{2014}.
\newblock \bibinfo{title}{Performance {Evaluation}}, in:
  \bibinfo{booktitle}{the CFA Institute Investment Foundations}, pp.
  \bibinfo{pages}{173--205}.
%Type = Book
\bibitem[{Conti(1978)}]{conti1978findings}
\bibinfo{author}{Conti, D.M.}, \bibinfo{year}{1978}.
\newblock \bibinfo{title}{Findings of the {Standard Benchmark Library Study
  Group}}.
\newblock \bibinfo{number}{500-538}, \bibinfo{publisher}{Sept. of Commerce,
  National Bureau of Standards, Institute for Computer Sciences and
  Technology}.
%Type = Inproceedings
\bibitem[{Deng et~al.(2009)Deng, Dong, Socher, Li, Li and
  Fei-Fei}]{deng2009imagenet}
\bibinfo{author}{Deng, J.}, \bibinfo{author}{Dong, W.},
  \bibinfo{author}{Socher, R.}, \bibinfo{author}{Li, L.J.},
  \bibinfo{author}{Li, K.}, \bibinfo{author}{Fei-Fei, L.},
  \bibinfo{year}{2009}.
\newblock \bibinfo{title}{Imagenet: A large-scale hierarchical image database},
  in: \bibinfo{booktitle}{2009 IEEE conference on computer vision and pattern
  recognition}, \bibinfo{organization}{IEEE}. pp. \bibinfo{pages}{248--255}.
%Type = Inproceedings
\bibitem[{Huppler(2009)}]{huppler2009art}
\bibinfo{author}{Huppler, K.}, \bibinfo{year}{2009}.
\newblock \bibinfo{title}{The art of building a good benchmark}, in:
  \bibinfo{booktitle}{Technology Conference on Performance Evaluation and
  Benchmarking}, \bibinfo{organization}{Springer}. pp. \bibinfo{pages}{18--30}.
%Type = Techreport
\bibitem[{IOSCO(2013)}]{iosco2013finben}
\bibinfo{author}{IOSCO}, \bibinfo{year}{2013}.
\newblock \bibinfo{title}{Financial Benchmarks}.
\newblock \bibinfo{type}{Technical Report}.
%Type = Inproceedings
\bibitem[{Jiang et~al.(2021)Jiang, Gao, Tang, Wang, Xiong, Luo, Lan, Li and
  Zhan}]{jiang2021hpc}
\bibinfo{author}{Jiang, Z.}, \bibinfo{author}{Gao, W.}, \bibinfo{author}{Tang,
  F.}, \bibinfo{author}{Wang, L.}, \bibinfo{author}{Xiong, X.},
  \bibinfo{author}{Luo, C.}, \bibinfo{author}{Lan, C.}, \bibinfo{author}{Li,
  H.}, \bibinfo{author}{Zhan, J.}, \bibinfo{year}{2021}.
\newblock \bibinfo{title}{{HPC AI500 V2. 0: The Methodology, Tools, and Metrics
  for Benchmarking HPC AI Systems}}, in: \bibinfo{booktitle}{2021 IEEE
  International Conference on Cluster Computing (CLUSTER)},
  \bibinfo{organization}{IEEE}. pp. \bibinfo{pages}{47--58}.
%Type = Article
\bibitem[{Johnson et~al.(2016)Johnson, Pollard, Shen, Li-Wei, Feng, Ghassemi,
  Moody, Szolovits, Celi and Mark}]{johnson2016mimic}
\bibinfo{author}{Johnson, A.E.}, \bibinfo{author}{Pollard, T.J.},
  \bibinfo{author}{Shen, L.}, \bibinfo{author}{Li-Wei, H.L.},
  \bibinfo{author}{Feng, M.}, \bibinfo{author}{Ghassemi, M.},
  \bibinfo{author}{Moody, B.}, \bibinfo{author}{Szolovits, P.},
  \bibinfo{author}{Celi, L.A.}, \bibinfo{author}{Mark, R.G.},
  \bibinfo{year}{2016}.
\newblock \bibinfo{title}{{MIMIC-III}, a freely accessible critical care
  database}.
\newblock \bibinfo{journal}{Scientific data} \bibinfo{volume}{3},
  \bibinfo{pages}{1--9}.
%Type = Inproceedings
\bibitem[{Joslin(1965)}]{joslin1965evaluation}
\bibinfo{author}{Joslin, E.O.}, \bibinfo{year}{1965}.
\newblock \bibinfo{title}{Evaluation and performance of computers: application
  benchmarks: the key to meaningful computer evaluations}, in:
  \bibinfo{booktitle}{Proceedings of the 1965 20th national conference}, pp.
  \bibinfo{pages}{27--37}.
%Type = Article
\bibitem[{Kacker(2021)}]{kacker2021quantity}
\bibinfo{author}{Kacker, R.N.}, \bibinfo{year}{2021}.
\newblock \bibinfo{title}{On quantity, value, unit, and other terms in the jcgm
  international vocabulary of metrology}.
\newblock \bibinfo{journal}{Measurement Science and Technology}
  \bibinfo{volume}{32}, \bibinfo{pages}{125015}.
%Type = Inproceedings
\bibitem[{v.~Kistowski et~al.(2015)v.~Kistowski, Arnold, Huppler, Lange,
  Henning and Cao}]{v2015build}
\bibinfo{author}{v.~Kistowski, J.}, \bibinfo{author}{Arnold, J.A.},
  \bibinfo{author}{Huppler, K.}, \bibinfo{author}{Lange, K.D.},
  \bibinfo{author}{Henning, J.L.}, \bibinfo{author}{Cao, P.},
  \bibinfo{year}{2015}.
\newblock \bibinfo{title}{How to build a benchmark}, in:
  \bibinfo{booktitle}{Proceedings of the 6th ACM/SPEC International Conference
  on Performance Engineering}, pp. \bibinfo{pages}{333--336}.
%Type = Article
\bibitem[{Lewis and Crews(1985)}]{lewis1985evolution}
\bibinfo{author}{Lewis, B.C.}, \bibinfo{author}{Crews, A.E.},
  \bibinfo{year}{1985}.
\newblock \bibinfo{title}{The evolution of benchmarking as a computer
  performance evaluation technique}.
\newblock \bibinfo{journal}{MIS Quarterly} , \bibinfo{pages}{7--16}.
%Type = Inproceedings
\bibitem[{Liu et~al.(2021)Liu, Liu, Li, Fu, Yang, Song, Zhao, Wang, Peng, Chen
  et~al.}]{liu2021closing}
\bibinfo{author}{Liu, Y.}, \bibinfo{author}{Liu, X.}, \bibinfo{author}{Li, F.},
  \bibinfo{author}{Fu, H.}, \bibinfo{author}{Yang, Y.}, \bibinfo{author}{Song,
  J.}, \bibinfo{author}{Zhao, P.}, \bibinfo{author}{Wang, Z.},
  \bibinfo{author}{Peng, D.}, \bibinfo{author}{Chen, H.}, et~al.,
  \bibinfo{year}{2021}.
\newblock \bibinfo{title}{Closing the "quantum supremacy" gap: achieving
  real-time simulation of a random quantum circuit using a new sunway
  supercomputer}, in: \bibinfo{booktitle}{Proceedings of the International
  Conference for High Performance Computing, Networking, Storage and Analysis},
  pp. \bibinfo{pages}{1--12}.
%Type = Misc
\bibitem[{MIT()}]{automl2021git}
\bibinfo{author}{MIT}, .
\newblock \bibinfo{title}{Automl benchmark datasets}.
\newblock \bibinfo{howpublished}{[EB/OL]}.
\newblock
  \bibinfo{note}{\url{https://openml.github.io/automlbenchmark/benchmark_datasets.html}
  Accessed December 2, 2021}.
%Type = Inproceedings
\bibitem[{Pan et~al.(2005)Pan, Wang, Tung and Yang}]{pan2005finding}
\bibinfo{author}{Pan, F.}, \bibinfo{author}{Wang, W.}, \bibinfo{author}{Tung,
  A.K.}, \bibinfo{author}{Yang, J.}, \bibinfo{year}{2005}.
\newblock \bibinfo{title}{Finding representative set from massive data}, in:
  \bibinfo{booktitle}{Fifth IEEE International Conference on Data Mining
  (ICDM'05)}, \bibinfo{organization}{IEEE}. pp. \bibinfo{pages}{8--pp}.
%Type = Misc
\bibitem[{Statista()}]{app2021git}
\bibinfo{author}{Statista}, .
\newblock \bibinfo{title}{Number of apps available in leading app stores}.
\newblock \bibinfo{howpublished}{[EB/OL]}.
\newblock
  \bibinfo{note}{\url{https://www.statista.com/statistics/276623/number-of-apps-available-in-leading-app-stores/},
  accessed at Dec 2, 2021}.
%Type = Article
\bibitem[{Stevens et~al.(1946)}]{stevens1946theory}
\bibinfo{author}{Stevens, S.S.}, et~al., \bibinfo{year}{1946}.
\newblock \bibinfo{title}{On the theory of scales of measurement} .
%Type = Inproceedings
\bibitem[{Tang et~al.(2021)Tang, Gao, Zhan, Lan, Wen, Wang, Luo, Cao, Xiong,
  Jiang et~al.}]{tang2021aibench}
\bibinfo{author}{Tang, F.}, \bibinfo{author}{Gao, W.}, \bibinfo{author}{Zhan,
  J.}, \bibinfo{author}{Lan, C.}, \bibinfo{author}{Wen, X.},
  \bibinfo{author}{Wang, L.}, \bibinfo{author}{Luo, C.}, \bibinfo{author}{Cao,
  Z.}, \bibinfo{author}{Xiong, X.}, \bibinfo{author}{Jiang, Z.}, et~al.,
  \bibinfo{year}{2021}.
\newblock \bibinfo{title}{Aibench training: balanced industry-standard ai
  training benchmarking}, in: \bibinfo{booktitle}{2021 IEEE International
  Symposium on Performance Analysis of Systems and Software (ISPASS)},
  \bibinfo{organization}{IEEE}. pp. \bibinfo{pages}{24--35}.
%Type = Article
\bibitem[{Wang et~al.(2021)Wang, Xiong, Zhan, Gao, Wen, Kang and
  Tang}]{wang2021wpc}
\bibinfo{author}{Wang, L.}, \bibinfo{author}{Xiong, X.}, \bibinfo{author}{Zhan,
  J.}, \bibinfo{author}{Gao, W.}, \bibinfo{author}{Wen, X.},
  \bibinfo{author}{Kang, G.}, \bibinfo{author}{Tang, F.}, \bibinfo{year}{2021}.
\newblock \bibinfo{title}{Wpc: Whole-picture workload characterization across
  intermediate representation, isa, and microarchitecture}.
\newblock \bibinfo{journal}{IEEE Computer Architecture Letters} .
%Type = Techreport
\bibitem[{Wells et~al.(2016)Wells, Bland, Nichols, Hack, Foertter, Hagen,
  Maier, Ashfaq, Messer and Parete-Koon}]{wells2016announcing}
\bibinfo{author}{Wells, J.}, \bibinfo{author}{Bland, B.},
  \bibinfo{author}{Nichols, J.}, \bibinfo{author}{Hack, J.},
  \bibinfo{author}{Foertter, F.}, \bibinfo{author}{Hagen, G.},
  \bibinfo{author}{Maier, T.}, \bibinfo{author}{Ashfaq, M.},
  \bibinfo{author}{Messer, B.}, \bibinfo{author}{Parete-Koon, S.},
  \bibinfo{year}{2016}.
\newblock \bibinfo{title}{Announcing supercomputer summit}.
\newblock \bibinfo{type}{Technical Report}. Oak Ridge National Lab.(ORNL), Oak
  Ridge, TN (United States).
%Type = Incollection
\bibitem[{Zairi and Leonard(1996)}]{zairi1996origins}
\bibinfo{author}{Zairi, M.}, \bibinfo{author}{Leonard, P.},
  \bibinfo{year}{1996}.
\newblock \bibinfo{title}{Origins of benchmarking and its meaning}, in:
  \bibinfo{booktitle}{Practical Benchmarking: The Complete Guide}.
  \bibinfo{publisher}{Springer}, pp. \bibinfo{pages}{22--27}.

\end{thebibliography}


\bio{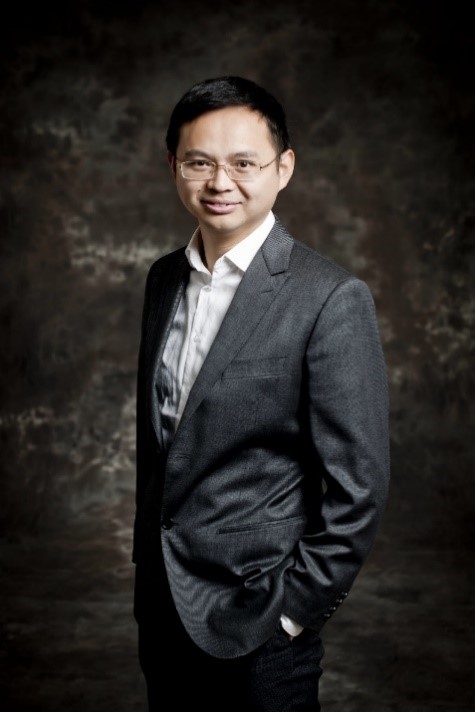}
Dr. Jianfeng Zhan is a Full Professor at Institute of Computing Technology (ICT), Chinese Academy of Sciences (CAS), and University of Chinese Academy of Sciences (UCAS), and director of the Software Systems Labs, ICT, CAS.  He received his B.E. in Civil Engineering and MSc in Solid Mechanics from Southwest Jiaotong University in 1996 and 1999, and his Ph.D. in Computer Science from Institute of Software, CAS, and UCAS in 2002. His research areas span from Chips, Systems to Benchmarks. A common thread is benchmarking, designing, implementing, and optimizing a diversity of systems. He has made substantial and effective efforts to transfer his academic research into advanced technology to impact general-purpose production systems. Several technical innovations and research results, including 35 patents, from his team, have been adopted in benchmarks, operating systems, and cluster and cloud system software with direct contributions to advancing the parallel and distributed systems in China or even in the world. He has supervised over ninety graduate students, post-doctors, and engineers in the past two decades. 
Dr. Jianfeng Zhan founds and chairs BenchCouncil and serves as the Co-EIC of TBench with Prof. Tony Hey. He has served as IEEE TPDS Associate Editor since 2018. He received the second-class Chinese National Technology Promotion Prize in 2006, the Distinguished Achievement Award of the Chinese Academy of Sciences in 2005, and the IISWC Best paper award in 2013, respectively. \endbio
\end{document}